# The potential role of AI agents in transforming nuclear medicine research and cancer management in India

Rajat Vahistha[1*], Arif Gulzar[1], Parveen Kundu[2], Punit Sharma[3,4], Mark Brunstein[5,6,7], Viktor Vegh[1]

1. Australian Institute for Bioengineering and Nanotechnology, The University of Queensland, Brisbane, Australia
2. Rohtak Nuclear Medcare Imaging Therapy and Research Center, India
3. Apollo Multispeciality Hospital, Kolkata, India
4. Apollo Hospitals Educational and Research Foundation, India
5. The University of Queensland, Brisbane, Australia
6. Australian eHealth Research Centre, CSIRO, Australia
7. Georgia Institute of Technology, Atlanta, United States

*Corresponding author

**Abstract**

India faces a significant cancer burden, with an incidence-to-mortality ratio indicating that nearly three out of five individuals diagnosed with cancer succumb to the disease. While the limitations of physical healthcare infrastructure are widely acknowledged as a primary challenge, concerted efforts by government and healthcare agencies are underway to mitigate these constraints. However, given the country's vast geography and high population density, it is imperative to explore alternative soft infrastructure solutions to complement existing frameworks. Artificial Intelligence agents are increasingly transforming problem-solving approaches across various domains, with their application in medicine proving particularly transformative. In this perspective, we examine the potential role of AI agents in advancing nuclear medicine for cancer research, diagnosis, and management in India. We begin with a brief overview of AI agents and their capabilities, followed by a proposed agent-based ecosystem that can address prevailing sustainability challenges in India's nuclear medicine.

1. Introduction

India's with population of 1.4 billion faces a significant cancer burden, with ~1.5 million new cases and ~850,000 deaths annually [1] [2]. With an incidence-to-mortality percentage of approximately 64.8%, nearly three out of five individuals diagnosed with cancer are expected to succumb to the disease [2]. Projections indicate that mortality rates will rise significantly, increasing from 64.7% to 109.6% between 2022 and 2050, largely due to demographic shifts as the reproductive-age population transitions into middle and old age. This growing cancer burden will place even more pressure on the already overburdened healthcare system, making it essential to address the gaps in both infrastructure and indigenous research and innovations to ensure timely and effective patient treatment [3].

This trend underscores the urgent need for a resilient, patient-centred framework that integrates medical advancements, early detection through diagnostics, timely therapeutic interventions, and equitable access to care. Nuclear medicine uses a small amount of targeted radioactive material to diagnose and treat cancer [4]. It plays a pivotal role in reducing cancer mortality by facilitating early detection through advanced imaging modalities such as Positron Emission Tomography-Computed Tomography (PET-CT). Utilizing radiopharmaceuticals, diagnostics contribute to approximately two-thirds of the broader nuclear medicine ecosystem, with the remaining one-third increasingly focused on targeted therapies [5].

The International Atomic Energy Agency (IAEA) highlighted that scaling up access to nuclear medicine and medical imaging services would avert nearly 2.5 million cancer deaths worldwide by 2030 and yield global lifetime productivity gains of USD 1.41 trillion – a net return of over USD 200 per USD 1 invested [4]. With the recent rapid growth of combining diagnostic imaging with targeted therapy using radioactive substances - radiotheranostics, therapeutic radiopharmaceuticals are improving survival and treatment outcomes, specifically for prostate cancer and neuroendocrine tumors [6]. However, indigenous production capacity is limited with only one public sector medical reactor producing therapeutic radionuclides (like $^{131}I$, $^{177}Lu$, $^{153}Sm$, etc.) in India. For most of therapeutic radiopharmaceuticals and many diagnostic radioisotopes, India's dependence on imported products raises costs and exacerbates logistical delays, further hindering timely cancer care [3]. In addition, India has only 30 to 50 medical cyclotrons, averaging one cyclotron every 28 to 46 million people [7]. The limited allocation of cyclotron facilities decreases access to many crucial radionuclides (18F, 64Cu, 89Zr, 124I and 68Ga etc.), leading to delays in cancer diagnosis and treatment, and hinders research.

Similarly, India has approximately 150 to 200 PET/CT scanners, which translates to roughly 1 scanner for every 7 to 10 million people. Many patients, particularly in rural and underserved areas, face challenges in accessing these vital diagnostic tools [3]. As of January 2025, only 720 Nuclear medicine facilities were licensed by the Atomic Energy Regulatory Board (AERB) of India. Despite rapid growth of PET-CT scanners, coverage is skewed with most of them concentrated in metropolitan areas and Tier II cities. Investments in physical infrastructure for nuclear medicine are crucial for improving coverage and reducing wait times. While, the limitations of physical infrastructures are widely recognized as a primary challenge, committed efforts by responsible agencies are underway to address this issue.

Despite India's global prominence in IT services, a significant gap remains in software infrastructure.. Artificial Intelligence (AI) agents are fundamentally reshaping the way complex challenges are addressed across various industries, with their impact on medicine being particularly transformative [8]. Recent global conferences, such as Microsoft Build, Meta Connect, and NVIDIA CES, have highlighted significant advancements in AI agents specifically tailored for healthcare applications. AI agents are evolving beyond being mere tools, and are now functioning as intelligent collaborators with the potential to shift healthcare from a reactive model to a predictive and personalized one [9]. The enthusiasm and innovation demonstrated at these conferences reflect a pivotal transformation in medicine, where AI agents not only enhance human expertise but also open new avenues for clinical care and research.

As developed nations drive these innovations, it is increasingly critical for developing countries, such as India, to adapt to the technologies. In this paper we consider the role of AI agents in improving the development and management of nuclear medicine, particularly for cancer. We first briefly describe AI agents and then provide a roadmap for how these can be deployed to address the prevailing challenges and improve overall efficiency and effectiveness.

## 2. AI Agents

AI agents are computer applications designed to achieve specific objectives by observing and interacting with their environment [10]. AI agent ability to act independently of human prompting enhance efficiency by automating routine tasks, analysing large datasets, and navigating hypothesis spaces with greater precision. Unlike traditional machine learning models (that typically require specialized training for each task and do not possess reasoning and interactive capabilities), AI agents are designed to adapt to new insights, incorporating the latest findings and refining hypotheses based on results [9]. This adaptability allows AI agents to stay relevant amid rapidly evolving scientific data, ensuring they balance new knowledge with existing information. A crucial component in training an AI agent is the availability of a human curated database. By utilizing curated databases, which provide structured and reliable data, these agents can mitigate the risks of misinformation often caused by hallucinations.

Figure 1 outlines the core components of AI agents: the model, external tools, and the orchestration layer. A highly advanced generative model, such as a large language model (LLM), processes incoming information, understands context, and generates appropriate responses based on its training and design. External tools complement the model by allowing the AI agent to interact with the real world. These include systems like APIs, databases, or other software that provide the agent with information or capabilities beyond its internal knowledge, such as retrieving live data, performing calculations, or executing specific tasks. The orchestration layer acts as the coordinator of the agent's operations. It manages how the agent receives and processes information, reasons through problems, and decides on the best course of action. This layer ensures that all parts of the agent work together smoothly, creating a unified and effective approach to completing tasks. Finally, the cognitive architecture ties these components together, giving the AI agent the ability to handle complex tasks.

For scientific explorations and assistance, AI agents are classified into four levels based on their level autonomy [9]. Human led agents at Level 0 have no autonomy, and scientists define and complete tasks with minimal AI assistance. Assistant agents at Level 1 formulate simple hypotheses and execute tasks defined by scientists. Collaborative agents at Level 2, generate hypotheses based on data trends and design rigorous experimental protocols, working closely with scientists. Finally, expert agents at Level 3 function as independent scientists, creating novel hypotheses, developing new experimental methods, and refining approaches based on results, with joint hypothesis formation and task completion alongside human scientists.

Key modules necessary for AI agents to interact with humans and experimental environments are perception, interaction, memory, and reasoning. Perception Modules are critical for agents to process diverse types of data beyond just text (multimodal data). These modules are responsible for converting and analysing various data forms, such as images, videos, and genomic profiles. Interaction Modules facilitate natural language communication between scientists and AI agents. These modules utilize chat interfaces that retain conversational history, allowing researchers to interact with agents in their own language. Memory Modules are divided into two types: long-term and short-term memory. Long-term memory stores critical knowledge from various sources, such as scientific literature and databases, and is continually updated with new or revised information. Short-term memory retains recent information, ensuring that more relevant and timely data is prioritized over older knowledge. This allows agents to inform future experiments and decisions based on past experiences.

Finally, Reasoning Modules are fundamental for enabling AI agents to assist in scientific tasks such as hypothesis generation, experimental design, and result interpretation. Reasoning can be implemented via prompt engineering or fine-tuning of LLMs, which empowers agents to plan experiments, evaluate hypotheses, and resolve competing mechanisms. The reasoning process can be single-path reasoning, which involves sequential task breakdown, or multi-path reasoning, which explores several alternatives before finalizing a plan. External feedback from scientists or the environment helps adjust agent actions, while internal feedback allows agents to self-evaluate and improve their decision-making. Detailed insights on how to build an AI agent for the different facets of biological data can be found at [9]. However, it is crucial to first understand the other associated challenges within the chain of events, before establishing intentional use cases of AI agents for nuclear medicine in India.

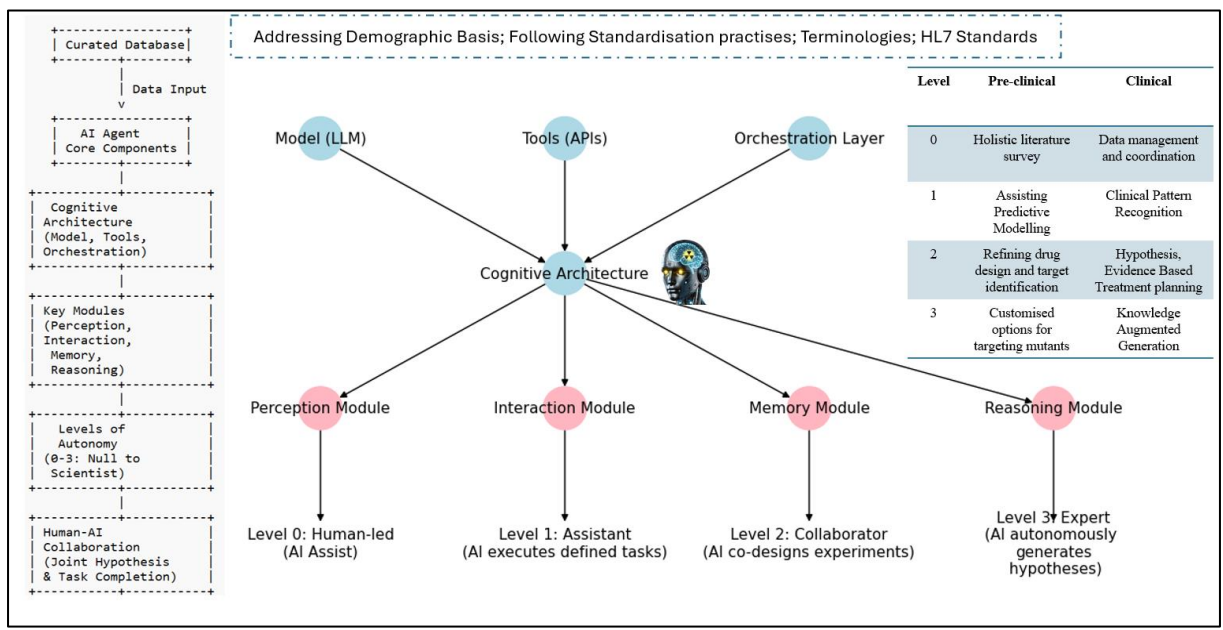

**Figure 1: Illustration of the cognitive AI agent (build using a large language model, interconnected with API's and orchestrated with different layers) with the necessary modules working at different level of autonomy.**

## 3. Challenges

Nuclear medicine faces several sustainability barriers [11], particularly in India and other developing nations, requiring a balanced approach across five different pillars as shown in Figure 2a. Human sustainability hinges on addressing chronic workforce shortages due to unavailability of a sufficient skilled and trained workforce and fostering diversity and inclusion to build a resilient and representative workforce.

Economically, mitigating the impact of supply disruptions through government and private entity joint ventures for reactor and cyclotron installation; and exploring local innovations options for radioisotopes and cold kits. Ecologically, nuclear medicine's soft infrastructure resists innovation, like AI-driven protocol optimization, especially in limited resource settings with high volumes of patients. Environmentally, minimizing the carbon footprint by optimizing energy consumption, reducing unnecessary procedures, and considering carbon offsetting strategies is essential. Socially, ensuring equitable access to diverse populations based on race, ethnicity and vast geographical locations is also essential. Figure 2b outlines these challenges from the perspective of beneficiaries at various technology readiness level (TRL, 1-9), varying from pre-clinical research to delivering a novel innovation to patients at clinics. We have outlined these challenges from the perspectives of preclinical and clinical research, along with a use case scenario, further extending the discussion to their translation into clinical trials.

In a diverse country like India, where genetic, cultural, and environmental factors vary significantly across regions, it is crucial to integrate demographic diversity into clinical data, as illustrated in Figure 2c (insights on diversity on data are provided in supplementary material). Computational platforms (access to high performance computing infrastructure), limited specialized training programs resulting in a shortage of skilled professionals, and inadequate collaboration among academic, research and clinical institution prevents holistic cancer management. Lack of data repositories is a widespread problem. Even when datasets are available, they may not represent the diverse genetic, environmental, and lifestyle factors that influence cancer in India. This can lead to biased AI predictions.

Limited preclinical research in India is hampered by economical and human sustainability challenges which divert funds from research. The development of targeted radiotracers is essential for advancing nuclear imaging and therapy, particularly for emerging isotopes like Cu-64 (PET imaging), Zr-89 (long-term immuno-PET), and Pb-212 (alpha therapy). However, in India, preclinical tracer development is hindered by lack of specialized synthesis facilities. Also, radiochemistry labs capable of handling complex ligand design, chelation chemistry, and biological validation are sparse. Preclinical radioactive imaging facilities for small animals (PET/SPECT/CT) are limited, reducing the efficiency of in vivo tracer validation. The approval process for novel tracers is slow, impacting their translation into clinical trials.

The substantial burden of primary care responsibilities borne by clinicians frequently results in burnout, leaving them with limited capacity to engage in research and the adoption of new technologies. The translation of research into clinical practice is further hindered by time and resource constraints faced by clinicians, along with economic barriers such as the high cost of technology and radiopharmaceuticals, which are exacerbated by the limited research conducted within the broader community rather than in medical centres.

Ensuring clinical efficiency requires seamless data sharing among nuclear medicine institutions; however, this remains limited due to several challenges, including: 1) varying data acquisition protocols among institutions creating silos of data that impede seamless data exchange and hinder research and collaboration; 2) the diversity of systems and software used across nuclear medicine facilities, with many centres relying on outdated, proprietary systems that lack interoperability; and 3) the absence of standardized terminologies and protocols for data sharing specific to nuclear medicine, which further complicates data reporting and integration. Limited data sharing is particularly critical for developing patient centric care for social sustainability, particularly for patients migrating from rural areas (where initial diagnosis has been done) to advanced clinical settings (for specialized treatment).

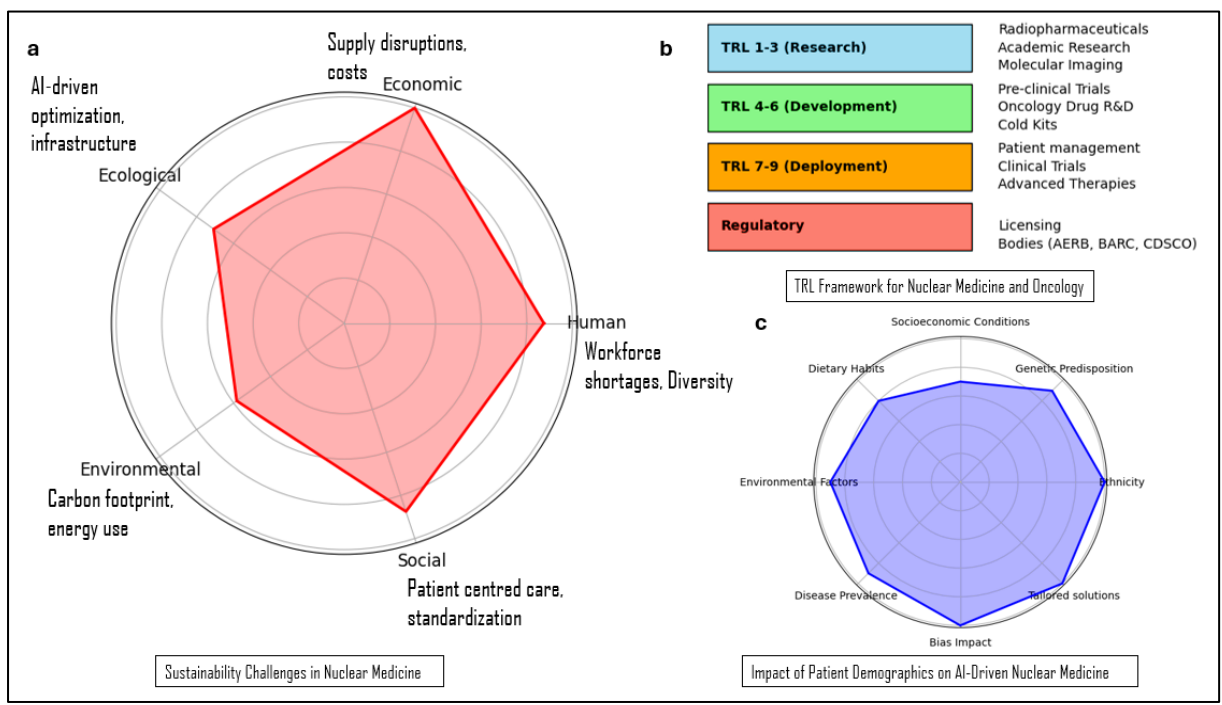

**Figure 2:** a) Identifying and measuring the extent of the sustainability challenges for nuclear medicine, b) Representation of stakeholders at different technology readiness levels, c) Dependability of AI on the patient demographic.

Conducting clinical trials in India faces a mixture of these challenges as well as workforce shortages, economic instability, and the prioritization of existing services over research. Rapid technological advancements place pressure on resources, requiring constant updates to trial protocols. Additionally, building trust in clinical trials is essential, and this requires confidence in the data which is hindered by the lack of standardized implementation for seamless data exchange among healthcare systems, research institutions, and laboratories. Recursive and time-consuming regulatory compliance and requirements are another factor that needs to optimise, as they ensure patient safety, data privacy, and adherence to standards. Economic instability, exacerbated by supply disruptions and fluctuating costs of essential materials such as radioisotopes and equipment, introduces uncertainty in trial planning and budgeting, affecting feasibility. Figure 3 illustrates how AI agents can contribute to a holistic approach to cancer research, diagnosis, and management in India, addressing various challenges across different domains. Sections 4 and 5 delve deeper into the roles of AI agents at both primary research levels (TRL 1-5) and clinical implementation stages (TRL 6-9), highlighting how these agents, with varying levels of autonomy, can effectively tackle research and infrastructure challenges for cancer using nuclear medicine. In section 6, we have outlined the opportunity in developing the agent-based approaches to population conditional outcomes while addressing the present challenges.

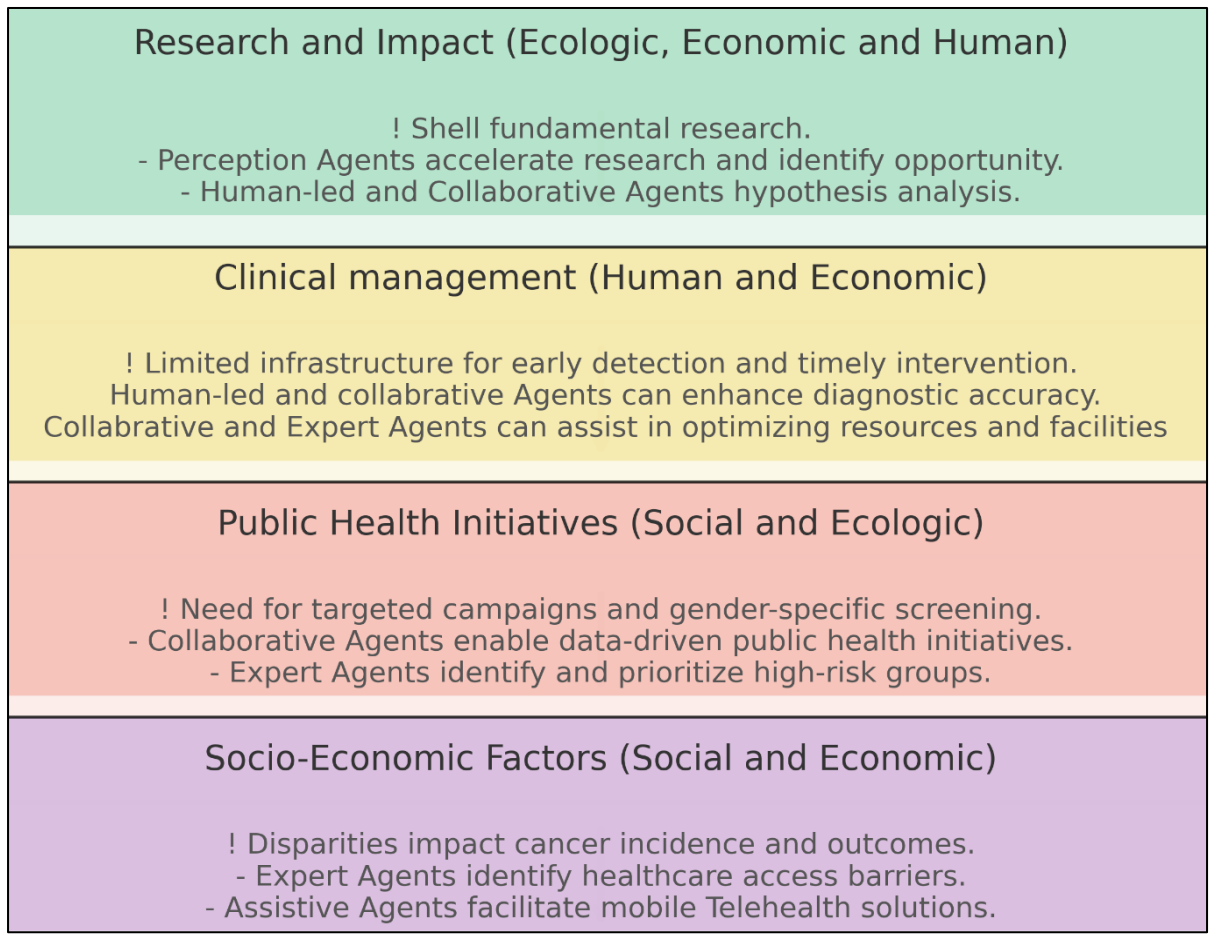

**Figure 3: Illustration of key challenges for cancer research and disease management in India.**

## 4. Agents for translational nuclear medicine research

Preclinical research for nuclear medicine in India is largely restricted to a few facilities, with a primary focus on the application of existing technologies rather than the development of novel innovation. In recent years, 67 radiopharmaceuticals were FDA approved, with 54 used for disease diagnosis and only 13 for therapeutic applications [12]. To best of the knowledge, none has been from indigenous research, despite the fact, India has a longstanding history in the field of radioisotopes, dating back to the visionary efforts of Dr. Homi J. Bhabha, which led to early production and utilization of radioisotopes in the country [8][1]. It suggests that the fundamental and translational research ecosystem remains stagnant, requiring urgent and proactive intervention. To highlight the potential of AI agents in revitalizing fundamental research and its impact, the following use cases have been explored.

Human-led AI agents have the potential to assist in knowledge discovery, particularly essential for developing new radiopharmaceuticals. These agents can optimise the practise by analysing vast chemical libraries like PubChem to identify promising new small molecules for synthesis and testing. AI has shown promise in predicting the binding affinity of a radiopharmaceutical to its target, increasing the likelihood of favourable pharmacokinetics and minimizing non-specific binding. This can be exemplified by the development of radioligands targeting prostate-specific membrane antigen and chemokine receptor-4 using structure-based design modelling [9]. In terms of human sustainability, human led (with perception modules) agents can also assist in streamlining routine tasks, enabling experienced radiochemists to focus on higher-level activities, effectively mitigating workforce shortages.

AI assistants (with additional interaction modules), can autonomously run the predictive models and design simple experiments, enabling quicker hypothesis testing and accelerating the pace of research [13]. For example, these agents can contribute to personalised preclinical dosimetry for radioligand therapies, a critical aspect in evaluating the safety and efficacy of these treatments. Monte Carlo simulations, the conventional method for dosimetry, are grounded on the selected population designed phantoms, computationally intensive and time-consuming [14]. Agents can be deployed to predict the results of Monte Carlo simulations with high accuracy, significantly reducing computation time. This allows for faster and more efficient evaluation of dose distribution in preclinical models. Addressing the challenge of human sustainability by reducing manual work, and economic sustainability by automating time-consuming tasks, helps optimize resource allocation in an environment where financial constraints are prevalent.

Collaborative and expert agents work in partnership (aided by reasoning modules) with scientists to refine hypotheses and design experiments that tackle more complex problems, such as radio-drug design and target identification. Recently, in silico design using AI to de novo binders has been used for CAR-T, showing potential for targeting mutated proteins that are resistant to clinically used antibodies [15]. In addition, a multi-agent system built on Gemini 2.0, helps uncover new, original knowledge and formulate demonstrably novel research hypotheses and proposals [16]. To address the social sustainability in research, these agents have the potential to analyse patient-specific data, such as tumor characteristics and immune profiles, to identify the most effective receptors.

Moreover, the low spatial resolution in preclinical PET is often a result of economic constraints. To reduce costs, larger crystals are used in the detectors, and the high expense of advanced scanners further limits the availability of higher-resolution imaging options. Image based expert AI agents have the potential to improve preclinical imaging by enhancing image resolution of scanner with smaller crystals using advanced techniques like sparse coding single image super-resolution and by improving attenuation correction using generative models. This can directly address economic sustainability and help optimize resources for research [17].

5. **Agents for clinical efficiency and management**

In clinical practice and management for cancer, one of the primary issues is the lack of data standardization. Without a unified approach to data coding and documentation, inconsistencies arise, making it difficult to aggregate and analyse data effectively. The use of standards in clinical practice (such as HL7® and openEHR), can help overcome this health infrastructure challenge and offer benefits to clinicians, including improved interoperability, enhanced clinical decision support, and streamlined administrative processes [18]. HL7's FHIR® (Fast Healthcare Interoperability Resources), a modern standard for exchanging healthcare information electronically, enables more seamless sharing of data. SNOMED CT (Systematized Nomenclature of Medicine - Clinical Terms) and LOINC® (Logical Observation Identifiers Names and Codes) enhance clinical decision support by providing a standardized and comprehensive language for medical concepts, lab tests, and observations [19]. These existing medical ontologies lack completeness in nuclear medicine concepts. NucLex, an extension of RadLex by the SNMMI AI Task Force, aims to address this gap by creating a controlled and publicly available ontology for standardizing nuclear medicine terms. While still in its early stages, similar efforts should be integrated into India's nuclear oncology ecosystem ensuring standardisation [18].

At the primary level of autonomy, human-led AI agents with perception modules can improve data integration by accurately interpreting standardized terminologies, addressing a key challenge. Figure 3a illustrates this with a patient presenting to his primary care physician (Dr A at Metropolitan Hospital) with complaints leading to a suspicion of prostate cancer. He is referred for imaging and genetic testing to diagnose and assess eligibility for an ongoing clinical trial at Cancer Research Centre. Following a FHIR-based workflow, a mpMRI scan was ordered and performed, with images and preliminary tumour measurements stored in the local PACS as an 'ImagingStudy' FHIR resource. The radiologist interpreted the scan, selecting key images for a report detailing tumour characteristics and recommending genetic profiling. After reviewing, the primary care physician confirmed trial eligibility and referred the patient to the Cancer Research Centre, sharing the mpMRI study via the standardised Metropolitan Health Information Exchange (Metro HIE) in compliance with guidelines [20].

This reduces the administrative burden on clinicians and researchers, allowing them to focus more on complex aspects of patient care and research. Helping addresses human sustainability by alleviating clinician burnout caused by repetitive administrative duties, is important, especially in clinics with high patient turnouts. In case of clinical trials, agents at this level might not be directly involved in the trial process but can aid in data management by automating basic tasks like organizing datasets and flagging inconsistencies [21]. For example, for the same patient in Figure 3A, including updated 'ImagingStudy' resources from follow-up scans and structured data entry for adverse events and outcomes, facilitates real-time monitoring while ensuring transparency. In addition, we have simulated an Indian origin registered trail at clinicaltrails.gov with ID, NCT05137561 to represent the potential of standardisation with AI agents (grey box).

> Register NCT-ID: NCT05137561
> Project title: Robotic-arm Assisted 68Ga PSMA PET/CT Guided Prostate Biopsy Versus MR-Directed TRUS Guided Prostate Biopsy clinical trial
>
> Problem: The principal investigator and his team wants to standardise real-time tracking of patient enrollment and procedures, establishing a transparent system for monitoring adverse events. Additionally, they recognize the need for efficient patient recruitment, as many eligible patients are not identified in time to enrol in clinical trials.
>
> Potential Solution: Research Team decide to implement open-source interoperability framework **openEHR and FEvIR** as core technologies in their workflow, with AI agents.
>
> Clinical trial-matching: Each patient is registered into the hospital's openEHR-based system. A mobile application based on FEvIR, looks at the patient's medical records and compares them to the requirements of available trials listed on clinicaltrials.gov. This helps doctors quickly see which trials a patient might be eligible for. There's also a version of this application that patients can use themselves, allowing them to learn about trials and express their interest in participating.
>
> Adverse events: One month into the trial, a participant, experiences an unexpected adverse reaction post-biopsy. Utilizing openEHR's structured reporting templates, the research team quickly logs the event. This allows for the automatic classification of the adverse event's severity based on predefined risk factors, notifying the principal investigator and ethics committee for timely intervention.
>
> Optimization: To optimize monitoring and decision-making, research team employs a three-tiered approach. Human-led decision support with agents at level 0 ensures that clinicians manually review real-time data and validate adverse event reports. Assistant and collaborative monitoring agents at level 1 and 2 to flag deviations in trial protocols, prompting clinician verification. Finally, fully autonomous expert agents' insights analyse procedural efficiency and patient outcomes, optimizing workflows for future trials.

At the next level of autonomy, as shown in Figure 4, AI agents begin to analyse (perception and interaction modules) clinical images, identify patterns, and assist in initial hypothesis generation based on available data [22]. At this level, agents can begin to address some of the challenges in economic sustainability, such as the high costs associated with diagnostic imaging and data processing, by optimizing resource usage and improving efficiency in clinical workflows. By assisting with early diagnosis and data analysis, these agents can also help clinicians make faster and more efficient decisions, addressing the workforce constraints in India's healthcare system. For example, agents integrating and analysing DICOM medical images with FHIR (DICOM on FHIR) holds great potential for enhancing interoperability between nuclear medicine imaging systems and other clinical systems [23].

In the context of clinical trials for a novel drug or therapy, AI agents at this level can facilitate the comprehensive integration of pre-treatment data across multiple modalities. For instance, whole-slide histopathology images, corroborated by PET/CT findings, and supplemented with genomic analysis when necessary for personalized assessment, can provide a holistic understanding of cancer [24]. Additionally, using AI agents to systematically analyse patient profiles and replicate these procedures for post-treatment evaluation can streamline the comparison of therapeutic efficacy, reducing redundancy and enhancing the efficiency of clinical trials. Comparison (pre- and post-) will aid trail sponsor/researcher for further explorations helping them with interpretation and hypothesis generation [25].

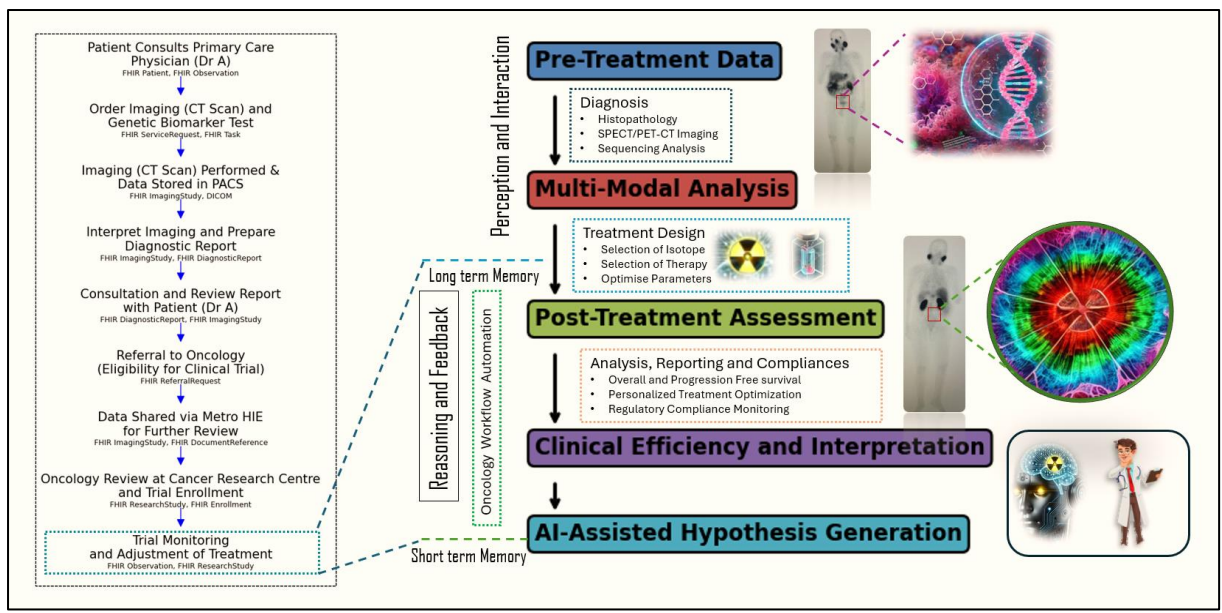

**Figure 4: a) Standardised FHIR resources are listed on the right. b) Holistic integration of the AI agent at the different level of clinical practices with varied level of autonomy**

At Level 2, AI agents transition to a more collaborative role, where they autonomously generate hypotheses based on data trends, design experiments, and refine existing protocols (utilizing memory modules). At this level, AI agents can help in patient referrals (involving FHIR tools) using automated screening based on eligibility criteria, significantly reducing time and resource constraints. Further, they can involve foundation models to learn from limited examples and then can be used to guide the segmentation of new images [24, 26]. This is useful for care centres involved with research practices. Moreover, Level 2 AI agents can assist with monitoring compliance during clinical trials by tracking patient adherence to protocols and real-time deviation identification, improving trial execution and ensuring that safety and regulatory requirements are met. This reduces the burden on clinicians and nodal agencies, who may struggle with time constraints due to economic barriers.

At Level 3, AI agents (involving perception, interaction, memory and reasoning modules) can predict patient responses to therapies based on genomic information and other medical data [27]. For example, in medical visual question answering, where AI agents interpret the question and provide an answer based on its analysis of the image and its associated textual data [28, 29]. For personalized decision support, theranostic digital twins, which serve as virtual representations of individual patients, offer a sophisticated approach to optimizing treatment strategies [30, 31]. Physiologically based pharmacokinetic modeling enables the integration of patient-specific data, encompassing both physiological and pathophysiological characteristics [31]. Model-based AI agents at this level can simulate various treatment scenarios, facilitating the optimization of critical parameters such as radiopharmaceutical selection, the number of therapeutic cycles, and the administered activity. This personalized approach moves beyond the limitations of the 'one-size-fits-all' paradigm, enhancing both clinical efficacy and social sustainability [32].

Further, these agents can drive significant advancements in clinical trial design by suggesting which endpoints are most likely to yield meaningful results based on historical and real-world data [33]. This level of autonomy addresses economic sustainability by optimizing resource allocation, which can otherwise lead to wasted resources if end points are poorly chosen. Furthermore, AI agents at this level can play a crucial role in translating preclinical findings into clinical trials by continuously refining patient recruitment strategies, ensuring that only the most suitable candidates are included in the trials, thereby improving the likelihood of successful outcomes [34]. By investing in AI-driven systems and fostering collaboration across sectors, India can overcome current barriers however, to develop the sustainable and innovative nuclear medicine ecosystem development challenges needs to be addressed.

## 6. Agents for public initiatives: opportunities and challenges

Public health initiatives face significant social and ecological sustainability challenges, such as the need for targeted campaigns and gender-specific screening programs [35]. Collaborative Agents play a vital role in enabling data-driven public health strategies by integrating diverse data sources, including epidemiological, demographic, and environmental factors. Expert Agents specialize in identifying and prioritizing high-risk groups by analysing health records, genetic predisposition, lifestyle factors, and imaging biomarkers. They can stratify populations based on their risk levels. For an example, we consider an Expert Agent deployed in a hospital network prioritizing women for personalized breast/cervical cancer screening. Women with a family history of such cancer with potential imaging markers are flagged for early and more frequent screenings, while those with low-risk profiles are recommended for standard screening intervals. The system also incorporates lifestyle and hormonal risk factors, such as prolonged estrogen exposure or obesity, to refine risk predictions. By prioritizing high-risk individuals for advanced diagnostic techniques like PET-CT scans, the Expert Agent ensures early detection, optimizing hospital resources.

Another use case for Expert Agent-based resource optimization is in radioisotope management. For example, a modern GE or IBA 18MeV 18F-producing cyclotron can yield up to 150 GBq per run, which is sufficient to provide 200–400 PET doses in a single 2-hour irradiation [36]. Similarly, installing advanced solid-target systems on exiting cyclotrons in India, such as IBA's Nirta solid target system and Comecer's Alceo 4.0 solid target system, and training professionals for production of radiometals such as Ga-68, Cu-64 , I-124and Zr-89 can achieve yields ranging from 10 GBq to several hundred GBq per run, translating into tens to hundreds of patient doses per production cycle. Thus, if fully optimized distribution networks were in place, even the limited number of cyclotrons could, per machine, generate enough activity to serve a much larger population, highlighting that the current challenge is not the physics of production but rather the infrastructure and geographic dispersion needed to provide timely access to these radionuclides

One major challenge is ensuring the robustness and reliability of these agents. AI systems often generate unreliable predictions, including hallucinations, reasoning errors, and systematic biases [37]. These issues are exacerbated by agent overconfidence in their flawed predictions and their sensitivity to the precise formulation of queries [18]. Additionally, the complexity of training multimodal systems and the computational demands of processing diverse data inputs pose significant hurdles. Effective error management strategies are crucial to maintaining system robustness and reliability, especially in interactive multi-agent systems where small issues can escalate into significant problems if not promptly addressed [38].

Another significant challenge is the evaluation and governance of AI agents. Developing comprehensive evaluation frameworks that consider ethical considerations, regulatory compliance, and the ability to integrate into discovery workflows is essential. The lack of standardization in biomedical discovery workflows and the variability in data generation protocols complicate the evaluation process. Furthermore, the governance of AI agents intersects technological, scientific, ethical, and regulatory domains, requiring robust guidelines to ensure responsible development and deployment. Addressing these challenges involves balancing innovation with accountability, establishing robust verification systems, and developing policies through initiatives to minimize regulatory gaps, as shown in Figure 5.

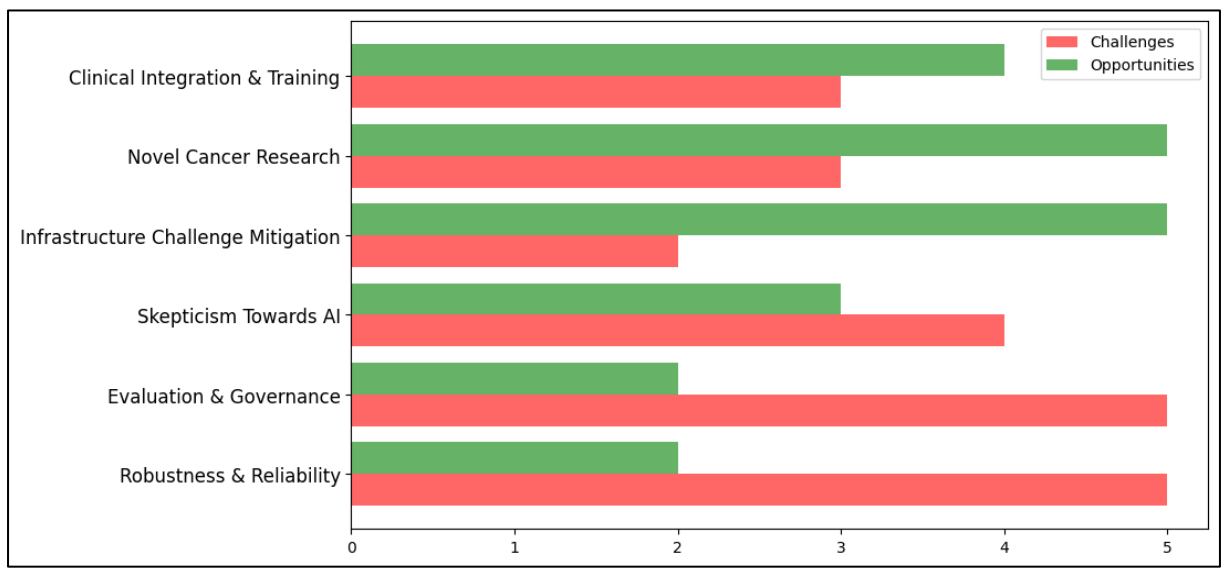

**Figure 5: Comparison of the wholesome challenges and likely opportunities for implementing AI agents.**

Recent discussions surrounding AI have shown growing scepticism about its abilities and ultimate value, particularly in fields like healthcare [38]. In the context of healthcare, AI has already demonstrated its transformative potential, particularly in fields like medicine and biology. For example, [39]. Despite the challenges faced by AI in other domains, healthcare benefits from decades of standardized medical knowledge, making it one of the more promising fields for AI applications. AI is being used to enhance diagnostic accuracy, optimize treatment plans, and support clinical decision-making. However, the success of these systems is heavily dependent on the quality of their design and training.

Several startups and researchers in India are using AI to enhance cancer diagnostics, imaging, and therapy monitoring, addressing the growing burden of cancer in the country. These innovators are utilizing advanced AI algorithms to improve early detection, precision medicine, and treatment outcomes, specifically targeting breast,, lung,and prostate cancers. In the academic and research space, private and government institutions are actively exploring AI applications in cancer imaging. However, despite the promising advancements, these efforts face the challenges addressed above (working is Silos), and opportunity lies in adapting quickly in designing the AI-agents (collective pool of efforts) to integrate seamlessly with existing hospital information systems, PACS, and laboratory systems, facilitating multi-modal data analysis and offering virtual training modules to upskill nuclear medicine physician, radiologists, oncologists, and technicians in using AI solutions effectively. Incorporating transparency and explainable features in agents helps clinicians understand AI predictions, thereby building trust and improving adoption. To explore collaborative decision-making processes, Human–AI interaction studies should be conducted. Such studies should be designed for AI agents' interaction with clinicians to evaluate the impact.

Regulatory bodies, along with concerned independent societies for nuclear medicine in India, should develop AI-specific guidelines to ensure compliance with data privacy, safety protocols, and regulatory standards. The European Union has ensured compliance for AI data sharing [40]. The US FDA has drafted artificial intelligence-enabled device software surveillance: lifecycle management and marketing submission recommendations that underscore the importance of applying a total product life cycle approach to AI-enabled devices, ensuring continuous oversight from development to post-market use [41]. A similar initiative is required to address the concerns to mitigate AI-specific risks for India. Devising AI agents for India's preclinical nuclear medicine ecosystem requires a structured, multi-phase approach integrating agentic modules, human, and institutional capabilities. Figure 6 emphasizes a modular approach (with details provided in the Supplementary sheet), integrating various capabilities to develop a robust and functional AI agents.

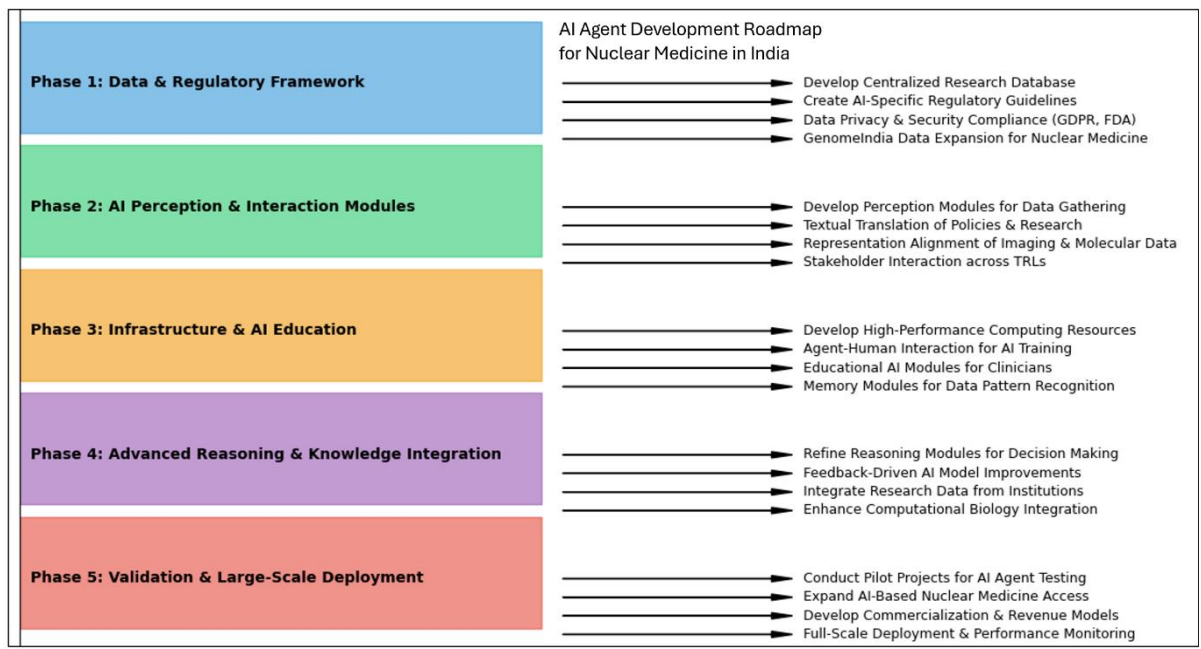

**Figure 6: Integrated five phase roadmaps to effectively apprehend the potential of the AI agents for nuclear medicine in India**

Despite the substantial structural and economical investment required to develop and deploy AI agents for nuclear medicine in India, the potential for reducing diagnostic times, streamlining clinical workflows, and fostering innovation far outweigh the upfront costs. Beyond direct cost savings, the AI agent's ability to accelerate research and development can generate long-term economic value. By facilitating faster drug discovery, enabling population-specific biomarker identification, and streamlining regulatory processes, it could reduce the R&D cycle. This also has the potential to open new revenue streams through technology exports, licensing, and public-private collaborations, making it a financially viable and transformative solution. Achieving this vision requires confronting challenges in India's nuclear medicine ecosystem. The journey demands collaboration, investment, and a shared vision.

**Acknowledgments**

Rajat Vashistha acknowledge the fellowship support from Cancer Council Queensland (CCQ) grant, aimed at Accelerating Collaborative Cancer Research (ACCR-190). We acknowledge the fruitful discussion with Saikat Ghosh on preclinical research in nuclear medicine.

**Supplementary:**

1. Challenges due to patient demographics

Patient demographics are a critical factor in the development and application of AI agents, as they directly influence the accuracy, generalizability, and fairness of AI-driven insights. Demographics such as age, gender, ethnicity, genetic predispositions, and socioeconomic conditions significantly impact disease manifestation, progression, and response to therapy. Figure 2c shows that in a diverse country like India, where genetic, cultural, and environmental factors vary widely across regions, it is essential to incorporate demographic diversity into the datasets used to train AI agents. Failure to do so can lead to biases in AI predictions, limiting their applicability to the Indian population.

In nuclear medicine, dataset repositories often originate from Western populations, where disease prevalence, radiotracer uptake, and tissue characteristics may differ significantly from Indian cohorts. For example, radiopharmaceutical pharmacokinetics may vary across populations due to genetic, dietary, and environmental factors. These differences, if unaccounted for, can lead to skewed AI predictions, affecting diagnostic accuracy and treatment efficacy for Indian patients.

Indian cohorts often exhibit distinct patterns of disease prevalence and presentation compared to global populations. For instance, certain cancers, such as lung, oral and cervical cancers, are more prevalent in India due to lifestyle and environmental factors. Similarly, metabolic diseases like diabetes often present differently in South Asian populations compared to Western populations, influenced by genetic predispositions and dietary habits.

Tailored datasets can not only address the social sustainability but also foster equitable healthcare by ensuring that diagnostic and therapeutic recommendations are specific to the local population. Incorporating Indian demographic data into AI agents also strengthens indigenous R&D capabilities. By analyzing population specific cohorts, researchers can identify population-specific biomarkers, optimize radiopharmaceutical protocols, and develop personalized therapies for Indian patients. This focus on local data reduces reliance on global datasets, fosters self-reliance, and ensures that innovations are relevant to the unique needs of the Indian healthcare system. Addressing these sustainability challenges is crucial for creating an environment that supports the development and validation of new therapies and diagnostic tools. In this context, we have explicitly outlined how the integration of AI agents can address these challenges and provide potential solutions.

2. Detailed Roadmap

A prototype for a centralized Indian research database using healthcare standards and explicitly designed for nuclear medicine, should be developed to aggregate fragmented data from various sources. This foundational step will establish the framework for further AI-driven applications. For example, the cancer imaging archive (TCIA) is an open-access database operated by the University of Arkansas for medical sciences. It hosts de-identified PET-CT, MRI images, organized by cancer type or anatomical site, and stored in DICOM format. The archive also includes supporting data (genomics) to enhance research, adhering to HIPAA and NIH data-sharing policies.

Subsequently, at the initial phase of the developing ecosystem for agents, focus should be on developing perception modules. They will play a pivotal role in gathering and processing diverse data sources related to India's nuclear medicine landscape. Textual translation facilitates the conversion of policy documents, research publications, and clinical guidelines into formats comprehensible by large language models. Meanwhile, representation alignment enables the integration of preclinical imaging data, radiopharmaceutical details, and molecular data from disparate sources, ensuring their representations are harmonized within a text-centered analytical framework. Interaction modules are equally critical during this phase, enabling meaningful engagement with stakeholders across various technology readiness levels.

The focus of the second phase should be on infrastructure development and resource allocation, where interaction modules are instrumental in implementing tool-use capabilities. These capabilities enable the AI agent to access and utilize high-performance computing infrastructure, executing tasks related to data processing, model training, and simulation. At this stage, interaction modules can play a pivotal role in both the development and educational aspects. Agent-human interaction modules are crucial for educating radiation safety and technology officers about AI applications in nuclear medicine. Acting as an interactive tutor, the AI agent could provide engaging learning experiences by explaining fundamental AI libraries for image denoising and molecular binding prediction - skillsets that will be indispensable in time.

Developing memory modules during this phase—following the creation of centralized databases—can enhance analytical tasks in nuclear medicine. These modules would support identifying patterns and trends, predicting treatment outcomes, and aiding in the discovery of new radiopharmaceuticals, thereby optimizing data-driven research and clinical decision-making. Continuous learning is essential for the AI agent to remain updated and enhance its performance over time. Short-term memory modules would allow the agent to recall previous actions and outcomes, thereby learning from experience and improving its efficiency in similar future tasks. Collaboration among clinics and engineering research institutions is essential for the successful deployment of the AI agent. A negotiation strategy can be implemented to determine data access and usage rights, ensuring compliance with data security and privacy regulations.

As the AI agent matures, reasoning modules will become essential in enabling higher levels of autonomy. Reasoning with feedback mechanisms would be instrumental in refining these processes. The AI agent would benefit from feedback provided by nuclear medicine experts, particularly during the testing and validation phase at medical universities and research centers. This iterative process aligns with the emphasis on incorporating human expertise in the reasoning loop. Furthermore, the agent would continuously adapt its reasoning and decision-making based on the outcomes observed in real-world preclinical scenarios. Long-term memory modules would function as repositories of critical knowledge related to radiopharmaceutical research, incorporating the latest research findings and best practices from faculty expertise at Indian institutions. These modules would also integrate relevant biological pathways and molecular interactions from computational biology and ensure adherence to standardized clinical guidelines in nuclear medicine protocols and procedures.

The third phase focuses on validating the AI agent through extensive pilot projects and trials. By applying the agent across various use cases its effectiveness can be assessed in real-world scenarios. Geographic expansion will democratize access to advanced nuclear medicine technologies, making them available beyond metropolitan areas. A sustainable commercialization strategy should also be developed during this phase, with revenue models based on subscriptions, licensing fees, or usage-based charges. This will ensure the long-term viability of the AI platform. In the final phase, AI agents will be deployed at large scale, with integration into all major nuclear medicine institutions. Its impact on R&D efficiency, clinical outcomes, and cost optimization should be evaluated through comprehensive metrics and key performance indicators.